# Mapping electronic states of dual-parallel and symmetric zigzag grain boundaries of graphene on highly oriented pyrolytic graphite

*Jun Ma[1], Xiaying Li[1], Longjing Yin[1]，Wenxiao Wang[1], Qi Sun[1], Yu Yang[2]\*, Ping Zhang[2], Jiacai Nie[1],*

*Changmin Xiong[1], Ruifen Dou[1]\**

[1]Department of Physics, Beijing Normal University, Beijing, 100875, People's Republic of China

[2] Institute of Applied Physics and Computational Mathematics, Beijing 100088, People's Republic of China

Email: rfdou@bnu.edu.cn and yang_yu@iapcm.ac.cn

ABSTRACT: The grain boundaries (GBs) of a graphene surface were extensively studied because GBs with specific defect configurations result in the formation of new curved structures, which can be treated as new carbon allotropes. We studied the structures and electronic spectra of two periodic GBs in graphene on highly oriented pyrolytic graphite (HOPG) surfaces using scanning tunneling microscopy and spectroscopy (STM/S). Our results demonstrated that a GB consisting of dual parallel periodic dislocation cores of pentagonal–heptagonal (5–7) carbon rings gives rise to an enhanced localized state at 0.45 eV above the Dirac point in graphene surfaces, which is attributed to van Hove singularities (VHSs). Moreover, the energy positions of the localized states are varied between 0.40 and 0.47 eV depending on the site and eventually decayed to 0.36 eV. The variation of the energy positions is induced by two parallel GBs because of the higher electron density at the GB as a defect center and the lower electron away from the GB. Another periodic GB with a symmetric zigzag structural feature induced VHSs at − 0.038 and 0.12 eV near the Fermi level. Moreover, intervalley scattering was observed on both these GBs. This means that carrier concentration and thus, the conductance around the periodic GBs can be significantly enhanced by localized density of states. This finding suggests that graphene with a properly embedded ordered GB is promising for improving the performance of graphene-based electronic devices.

Online supplementary data available from http://www.sciencedirect.com

# 1. Introduction

Grain boundaries (GBs), the interfaces between the domains of materials with different crystallographic orientations, are ubiquitous in highly oriented pyrolytic graphite (HOPG) surfaces that are generated via polycrystalline growth [1-7]. Two decades ago, Albrecht *et al.* were the first to observe a coiled-like structure of GBs in HOPG using scanning tunneling microscopy (STM) [1]. Simonis *et al.* used STM and determined that the periodic structure of coiled-like GBs comprised dislocation cores consisting of pentagon- and heptagon-shaped carbon rings, and corresponding structural models were developed as well [3]. Subsequently, Červenka *et al.* systematically investigated the structural and electronic properties of GBs consisting of the periodic pentagonal–heptagonal carbon rings in HOPG using STM and scanning tunneling spectroscopy (STS) [5-7]. They proved that the GBs in the carbon honeycomb lattice break the translation symmetry and induce quasilocalized electron states near the Fermi energy neighboring the GBs. Moreover, they reported room-temperature ferromagnetism induced by topological point defects [7].

Recently, the structures and electronic properties of GBs on large-area polycrystalline graphene monolayers, which are grown on metal substrates such as Cu, Ni, and SiC substrates by chemical vapor deposition (CVD) and epitaxial growth, have been thoroughly studied using transmission electron microscopy (TEM) and STM [8-24]. In the experiments, both disordered aperiodic and ordered periodic GBs were clearly observed on the polycrystalline graphene monolayers. The disordered aperiodic GBs act as electron scattering centers and degrade the electrical conductivity of graphene, which should be avoided to the utmost extent [22]. In contrast, the ordered GBs are predicted to have peculiar properties such as a high mechanical strength, magnetism, and a well-defined transport gap. Specially, the ordered GBs consist of a periodic pentagonal–heptagonal pair resembling those found in HOPG, and their structures and properties have been thoroughly investigated [23]. Owing to the positions of carbon atoms and local structural features in the ordered GBs distinctly being resolved by the atomic resolution techniques of TEM and STM, accurate atomic structural models of these periodic GBs have been constructed using theoretical calculations. Recent STM/S studies show that the periodic pentagonal–heptagonal pair dislocations exhibit van Hove singularities (VHSs) in the localized density of states (DOS), [18-20] in good agreement with theoretical data. Therefore, the relationship between the detailed structural arrangements of GBs and the specific properties of graphene induced by the ordered GBs can be easily established. Although a growing number of experimental and theoretical results have been used to understand the structures, accurate models, and corresponding electronic and mechanical properties of the observed GBs in graphene, ordered GBs consisting of various periodic dislocation cores have yet to be thoroughly explored. Theoretically, graphene GBs are predicted to have distinct electronic, chemical, and mechanical properties that strongly depend on their atomic arrangement. Thus, different atomic arrangements in GBs result in unusual and typical electronic, chemical, and mechanical properties, which have not been reported until now and will facilitate a deeper understanding of the effect of GBs on pristine graphene.

In this paper, we investigate the structures and electronic spectra of two ordered GBs in the graphene layer on HOPG using STM/S. Our results demonstrate that one periodic GB consisting of dual parallel periodic dislocation cores of 5–7 carbon rings results in localized state sat 0.47 eV above the Dirac point (DP) for the graphene surfaces, which is attributed to VHSs. Our density functional theory (DFT) simulation for the dual parallel 5–7 GBs revealed a localized state at 0.45 eV above the Fermi level (see Supporting Materials Figure 1), in agreement with the STS result. More importantly, the energy positions of the localized states are found to oscillate between 0.40 and 0.47 eV depending on the site, and eventually decay to 0.36 eV. The oscillation of energy positions is induced by the two parallel GBs because of the higher electron density at GB as a defect center and a lower one away from the GB. The other periodic GB with a symmetric zigzag structural feature generates further VHSs at −0.038 and 0.12 eV near the Fermi level. We find that both the linear periodic defects can induce intervalley scattering, which proves that the periodic GBs act as electron scattering centers. This means that the carrier concentration and thus, the conductance around ordered GBs can be significantly enhanced by localized DOS. Our findings strongly suggest that properly mediated ordered GB scan be potentially used to improve the performance of graphene-based electronic devices.

## 2. Methods

Samples of HOPG of ZYG quality were purchased from NT-MDT. ZYG-quality HOPG samples with a mosaic spread of 3.5°–5° were used because they have a high population of GBs on the HOPG surface. The HOPG samples were cleaved using an adhesive tape in air and transferred to a low-temperature STM (from UNISOKU) working under ultrahigh-vacuum (UHV) conditions (less than $10^{-9}$ mbar) at liquid nitrogen temperature. The HOPG samples were heated to 500°C in UHV before STM/S experiments were performed. All the STM/S experiments were performed in the constant current mode with either mechanically formed Pt/Ir or electrochemically etched tungsten tips. Local STS spectra were recorded using a lock-in amplifier with a modulation at 918 Hz and a 30-mV amplitude. The DFT calculation details are presented in the Supplemental Material on theoretical calculations [24,25].

**3**. Results and discussion

Figure 1(a) shows a typical STM image of a GB on a HOPG surface. Here, the GB appears as two bright lines. A magnified STM image (Figure 1(b)) of the square area in Figure 1(a) reveals dual parallel GBs. The distance between the two parallel bright lines is ~1.21 nm. The two bright lines protrude from the graphite surface at a height of up to 0.35 nm, as shown by the profile line in Figure 1(c). The height of the dual parallel GB is almost independent of the applied bias voltage. The two crystalline domains are separated by the dual parallel GBs. The dual GBs are characterized by a misorientation angle $\vartheta = \vartheta_1 - \vartheta_2$, where $\vartheta_1$ and $\vartheta_2$ are the lattice orientations of the two domains with respect to a common reference vector, for which we abide by the definition of the relative rotation angle in the range of *0°< $\vartheta$ < 60°* (small-angle GBs correspond to $\vartheta$ close to 0° or 60°) [26]. The two insets in the lower panel of Figure 1(a) show that the dual parallel GBs separating the two domains are characterized by a misorientation angle $\vartheta_a$=21.0 ± 1° and exhibit ordered one dimensional (1D) superlattices. The periodicity $d_a$ of the two ordered GBs along the line defect is ~0.65 nm. We also performed fast Fourier transform (FFT) analysis of the graphene area containing GBs, as shown in Figure 1(d). The results show two outer sets of hexagonal patterns corresponding to the reciprocal lattices of the two graphene domains. The two hexagonal patterns are tilted from each other by 20°, which confirms the accuracy of the measured relative rotation angle ($\vartheta_a$) of the two domains based on the reference vector of the graphene monolayer. The six bright spots in the middle demonstrate a (√3×√3)*R*30° superstructure resulting from intervalley scattering of the free electrons off defects, which has been observed around point defects and step edges in graphite and graphene layers [5, 27-30] .

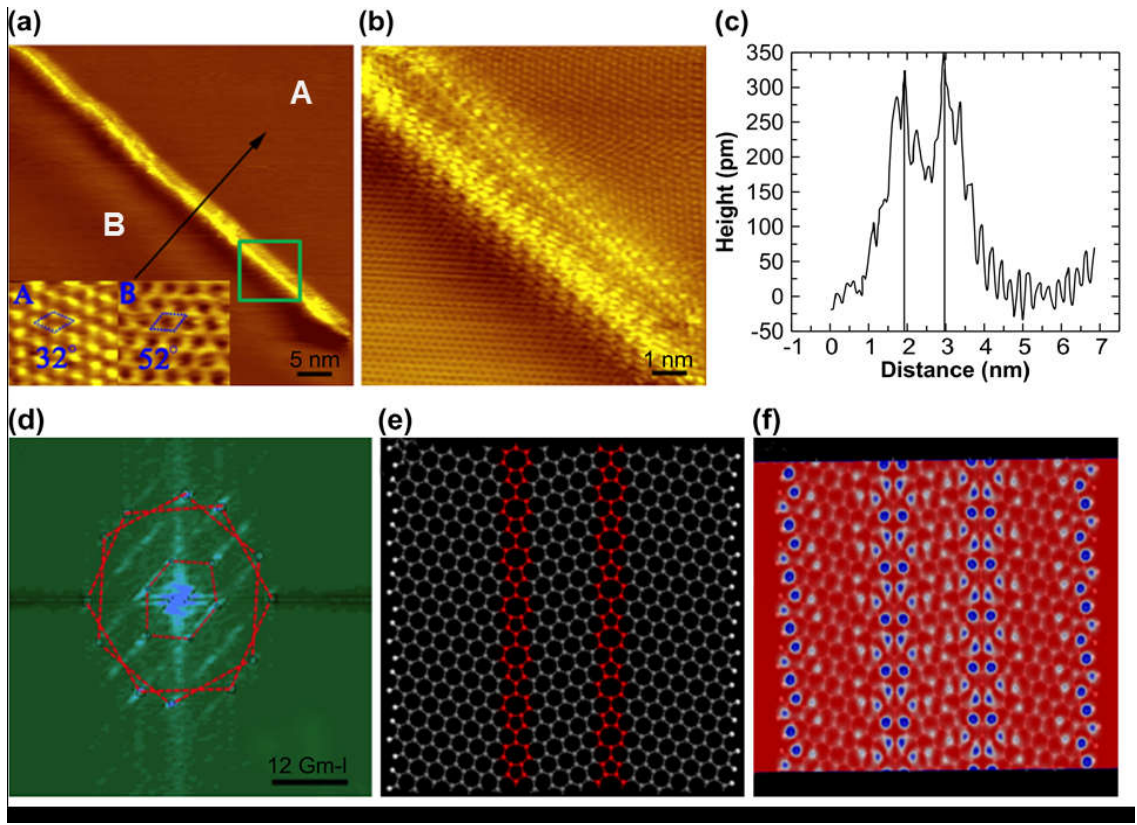

**Figure 1** (a) STM topography image of graphene layer with a bright line defect (GB) on HOPG ($V_{sample}$ = 0.52 V, I = 49.0 pA). (b) Magnified image of the green frame in (a) – an atomically resolved STM image of the dual parallel 5–7 GBs ($V_{sample}$ = 0.52 V, I = 49.0 pA). (c) Profile line along the black arrow in panel (a) showing the height of the line GB. (d) Fast Fourier transform of (a). The outer bright spots correspond to two sets of the reciprocal lattices of graphene. The domain misorientation angle between the two domains is 20°. The six spots in the middle are due to intervalley scattering. (e) Schematic of the dual GBs consisting of alternate 5–7 pairs formed by rotating two half-lattices in graphene by 21.8°. The pentagonal and heptagonal carbon rings are highlighted in red. (f) Simulated STM image obtained at 520 mV modeling the local DOS near the periodic 5–7 GBs, which corresponds well with the STM image of (b).

Periodic interference patterns of carbocyclic rings are also observed in the STM image along dual parallel GBs defects. These interference patterns, analogous to that resulting from the point defects in graphene, can be attributed to intervalley scattering between two adjacent Dirac cones in the first Brillouin zone of the graphene monolayer.[30] The inner spots arranged into two periodic lines in the FFT image represent the real-space periodicity of the 1D GB superlattices.

According to the structural features, i.e., the $\vartheta_a$ and $d_a$, of dual parallel GBs, such defects can be defined as two (1,0)|(0,1) line defects parallel to each other, which consist of alternate (5–7)carbon rings with a $d_a$ of 0.65 nm. Figure 1(e) shows a schematic of the dual parallel 5–7 GBs, supported by the coincidence site lattice theory proposed by Fasolina *et al.* in Ref. [31]. This optimized structural model of two (1,0)|(0,1) GBs reveals a misorientation angle of 21.0° between two grains, consistent with the experimental measurement. Even though a similar 5–7 periodic GB was detected in HOPG and a graphene layer using STM in the literature, [3, 14] here, we find that two 5–7 GBs form parallel line defects. Thus, we expect that the electronic properties of the dual parallel GBs will be different from those of the ordinary 5–7 GBs in graphene. Figure 1(f) shows a theoretically simulated STM image of a 46.40-Å-wide graphene sheet containing a dual parallel 5–7 GB obtained at a tip voltage of 520 mV. Except for some distributions at the saturating hydrogen atoms, one can see that the localized electronic states are mainly distributed around the dual parallel 5–7 GB. The simulated image is in good agreement with the measured STM result, confirming the fact that the localized electronic state near the Fermi level is introduced by the 5–7 GB.

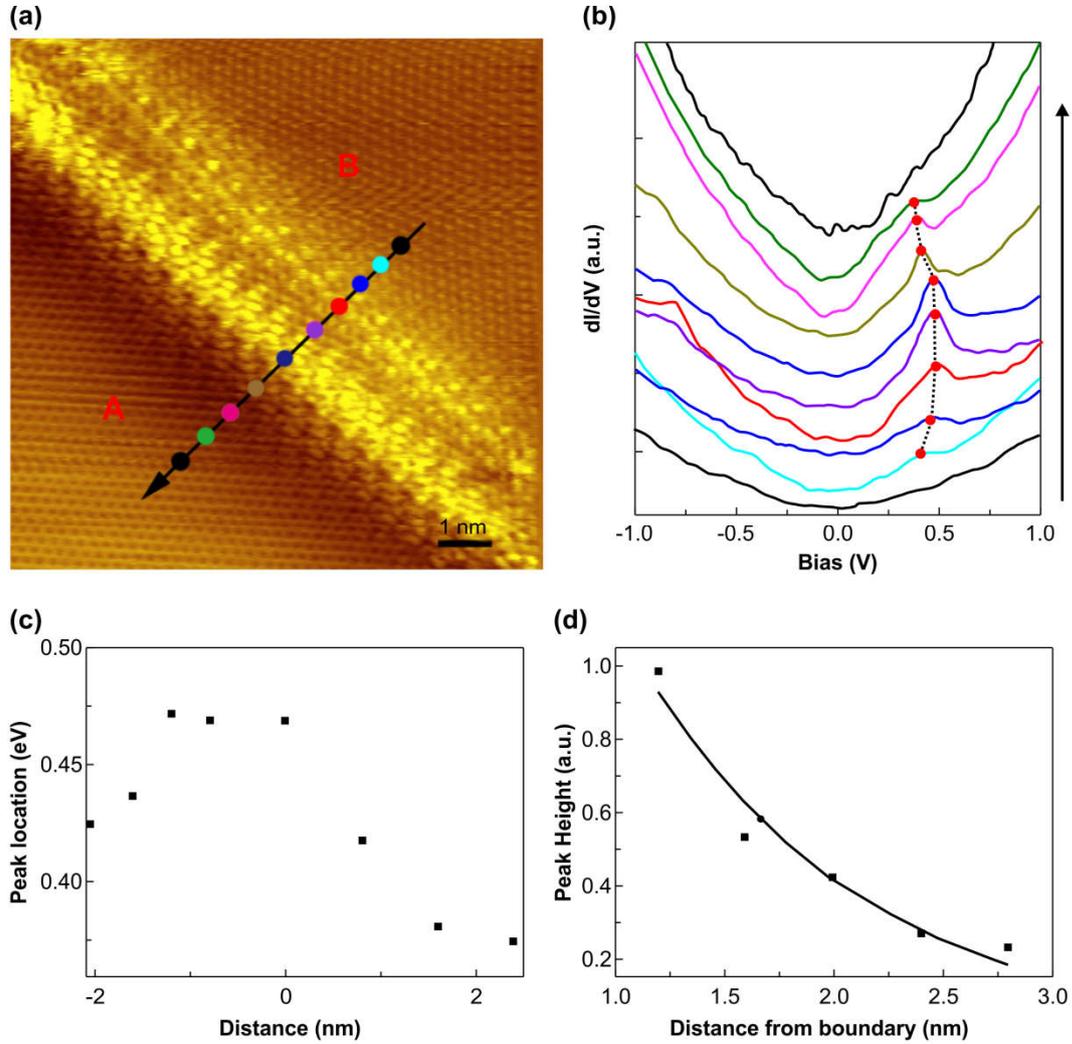

**Figure 2** (a) High-resolution STM image of dual parallel 5–7 GBs with a separation of 1.21 nm ($V_{sample}$ = 0.52 V, I = 49.1 pA). (b) Tunneling spectra acquired at colored spots along the black arrow from area A to B in (a). (c) Relationship between the energy position of distinct dI/dV peaks in (b) at ~400 mV and the corresponding spatial positions around the dual parallel GBs. The purple dot between the two boundaries in (a) is the zero position of the x-axis. (d) Peak height as a function of the localized state versus the distance from the dual GBs. The points can be fitted by $y = 0.033 + 0.887e^{-0.895x}$.

We also studied the electronic structure of the dual parallel GBs using STS. Figure 2(b) shows dI/dV spectra obtained at points across the dual parallel GBs, as indicated by the arrow from B to A in Figure 2(a). We note that the localized DOSs (obtained at the black, green, and pink dots) in area A are V-shaped and vanish at the DP, as expected for single layer graphene, and agree well with the honeycomb structure observed in the STM topographical images. Moreover, the DP is offset by ~ –0.08 eV below the Fermi energy due to electron doping caused by defects in the HOPG. In contrast, the localized DOSs (located at the black, light blue, dark blue, and red dots) in area B are parabolic curves that are characteristic of HOPG, indicating coupling of the top layer to the substrate supported by the triangle lattice, detected using STM. This means that the periodic dual parallel GBs separate the domains into the single layer graphene SLG (A) and the coupling carbon layer (B). More importantly, except for the two curves obtained at the black dots at a distance of 2.5 nm from the GB defects, all the STS spectra have distinct enhanced localized DOS peaks in the positive bias range. The localized DOSs at the red, purple, and black blue points are located at ~0.47 eV, which is in good agreement with the previous theoretical results [22]. We also calculated the DOS for the 46.40-Å-wide graphene sheet containing a dual parallel 5–7 GB and found a localized state at 0.45 eV (see Supplemental Material Figure 1). Accordingly, we deduce that the site-independent local DOSs very near the GBs are mostly induced by VHSs, which are the signatures of 1D states localized at the interface, as observed in previous studies [19,20,28] . Meanwhile, we found that the localized DOSs at sites 0.50–2.5 nm away from the GBs reduced from 0.42-0.40 eV to 0.38 eV. We derived a function relationship between the energy positions of the enhanced localized DOS peaks and the corresponding spatial distance from the dual GBs (here, the purple dot between the two boundaries in Figure 2a is set as the zero position of the X axis), as shown in Figure 2(c). The energy positions vary from 0.40 to 0.48 and finally decay to 0.38 eV with the graph being centered at 0 nm. The oscillation behavior at the energy positions is similar to that in the charging and discharging processes of the parallel-plate capacitor. Hence, dual parallel GBs with a small separation distance of 1.21 nm can be considered as two plate capacitors. During STS measurement, curves are obtained along the arrow (B→A) using a tip as a gate; the dual parallel GB, analogous to two plate capacitors, causes electron oscillation between two parallel GBs. It is shown that close to the GBs, 1D localized electron states appear from 0.40 eV up to–0.45 eV and far from the GBs, 1D localized states gradually decrease from 0.45 eV down to 0.40 eV. Thus, the oscillation behavior appears similar to the charging and discharging behavior of two parallel plates. We further note that the energy intensity (peak height) of different localized DOS peaks varies depending on the distance from the GBs. The relationship between the peak height and the distance from the dual GBs can be fitted by the exponential function $y = 0.033 + 0.887 e^{-0.895x}$. This means that the intensity of the localized DOSs is decreased at sites far away from the GBs.

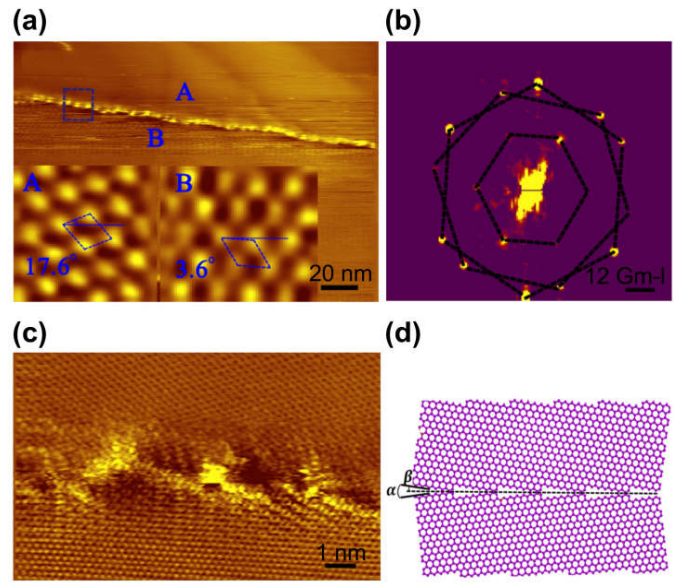

**Figure 3** (a) Topography of the GB on graphite (200 nm × 200 nm, $V_{sample}$ = –0.54 V, I = 59.0 pA). The insets show magnified atomic resolution images of areas A (1.14 × 1.14 nm) and B (1.12 × 1.12 nm), which show the relative orientation of the two domains separated by the GB. (b) Fast Fourier transform of (c) showing the periodicities of two sets of hexagonal lattices. The inner pattern is induced by the GB. The FFT indicates that the rotation angle between the two patterns is ~21°. (c) Magnification of the square in (a) with an obvious atomic resolution of the zigzag GB. The aromatic ring patterns are due to intervalley scattering ($V_{sample}$ = –0.48 V, I = 48.7 pA). (d) Schematic of the symmetric zigzag GB corresponding to image (c).

Figure 3(a) shows another type of ordered GB observed on HOPG. Figure 3(c) shows a magnified image of the square area in Figure 3(a), which reveals a bright line with a periodic and symmetric pattern. The insets in Figure 3(a) show two high-resolution STM images of areas A and B separated by the GB. These images indicate that the lattice misorientation angle of the two domains with respect to a common reference vector is ~21.2°. The two domains are thus separated by a GB with a misorientation angle $\vartheta_b$ =21.2±1° and a 1D periodicity $d_2$ of 1.28 nm. Figure 3(b) shows the FFT of the area in Figure 3(a), which is very similar to the FFT image in Figure 1(d). Here, the two sets of outer hexagonal patterns correspond to the reciprocal lattices of the two graphene domains. The two hexagonal patterns are tilted with respect to each other by 21°, which is in agreement with the measured misorientation angle ($\vartheta_b$ = 21.2°). The six bright spots in the middle indicate a (√3×√3) *R*30° superstructure attributed to the effect of intervalley scattering of the electrons. The periodic interference patterns of carbocyclic rings in the STM image are also observed along the dual parallel GB defects in Figure 3(c). The inner spots arranged in two periodic lines in the FFT image represent the real-space periodicity of the 1D GB superlattices.

Based on the rotation angle $\vartheta_b$ and periodicity $d_b$ of a GB, the above symmetric GB can be defined as two symmetric (7,1)|(1,7) zigzag defects with $d_b$ = 1.28 nm. A schematic of the zigzag GB is shown in Figure 3(d). The structure of the GB can be described using a simple model where the superlattice periodicity is determined by only two parameters: the misorientation angle (α) between two grains and the rotation angle (β) of a GB relative to the graphite lattice [31,32]. The superlattice periodicity $d_b$ is given by $d_b = \sqrt{3}d/2\sin\beta$, where β= α/2 and *d*= 0.247 nm is the graphite-lattice parameter. In our case, α= 20°; thus, $d_b$ = 1.23 nm, which is very similar to the experimental result. This GB structure is commonly observed in graphene; however, the electronic properties of the symmetric zigzag structure remain relatively unexplored.

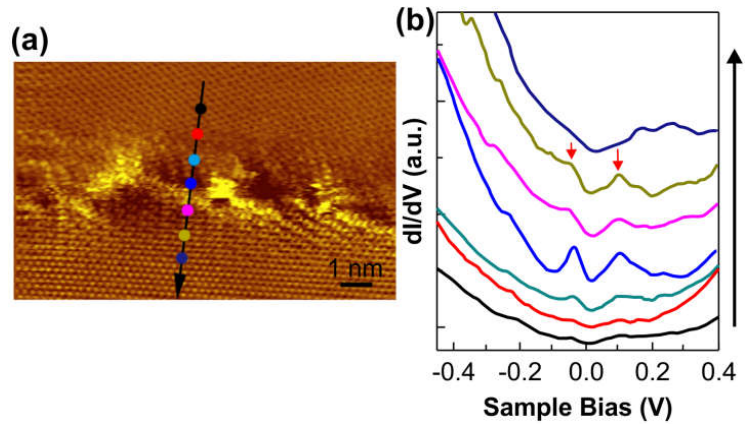

**Figure 4** (a) Topographic image with an obvious atomic resolution of the GB (b) d*I*/d*V* spectra obtained along the black arrow at positions corresponding to the dots in (a).

The experimental dI/dV curves acquired across the GB along the arrow (in Figure 4(a)) are shown in Figure 4(b). All the STS curves obtained in our experiments are V-shaped and the detected DP is located 0.05 eV below the Fermi level. This means that the area with a periodic GB exhibits features of single layer graphene. Two prominent peaks in the STS spectra at – 0.038 and 0.12 eV are observed at the GB and at a distance 0.5 nm away from the GB, respectively. In contrast to the features at the two parallel (1,0)|(0,1) 5–7 GBs, the energy positions of the localized DOS peaks are site-independent. These site-independent localized DOSs prove that the enhanced peaks in the STS spectra are induced by VHSs and not by the defect states in single layer graphene. Previous theoretical calculations predicted that the zigzag defects in graphene should cause a net magnetic moment; however the value of the net magnetic moment is inconsistent with each other in the theoretical calculations [33]. We calculated the local DOSs at a GB consisting of five (7,1)|(1,7) zigzag supercells with a periodicity of 1.23 nm (see Supplemental Material Figure 2). We found that the calculated energy positions of the enhanced local DOSs are not in good agreement with those in the experiment. In our calculation, the localized DOS at the symmetric zigzag defect is calculated by considering a antiferromagnetic moment of 0.1$\mu_B$, which might be the reason for the deviation between the calculated and the experimental results.

**4.** Conclusion

In conclusion, we studied the structures and electronic spectra of two periodic GBs in graphene on HOPG using STM/S. Our results show that one GB consisting of two parallel (1,0)|(0,1) dislocation cores of 5–7 carbon rings results in a localized VHS state at 0.45 eV above the DP. More importantly, these localized states are site dependent. This oscillation can be tuned using the dual parallel 5–7 GBs. The other periodic GB with a symmetric zigzag structural feature induces VHS states near the Fermi level. Intervalley scattering was observed on both these GBs, which means that the carrier concentration and thus, the conductance around the ordered GBs can be significantly enhanced by localized density states. This finding provides a rational method to improve the performance of graphene-based electronic devices by properly mediating and designing ordered GBs in a graphene layer.

**Acknowledgement**

This work was supported by the National Natural Science Foundation of China (Grant No. 11474024). This work was also supported by the open project of the State Key Laboratory of Low-Dimensional Quantum Physics, Tsinghua University.

**Conflict of Interest Disclosure:**

The authors have no competing interests to declare in this paper.

Network. *Phys. Rev. B* **2008**, *77*, 125425.

# A Graphical abstract

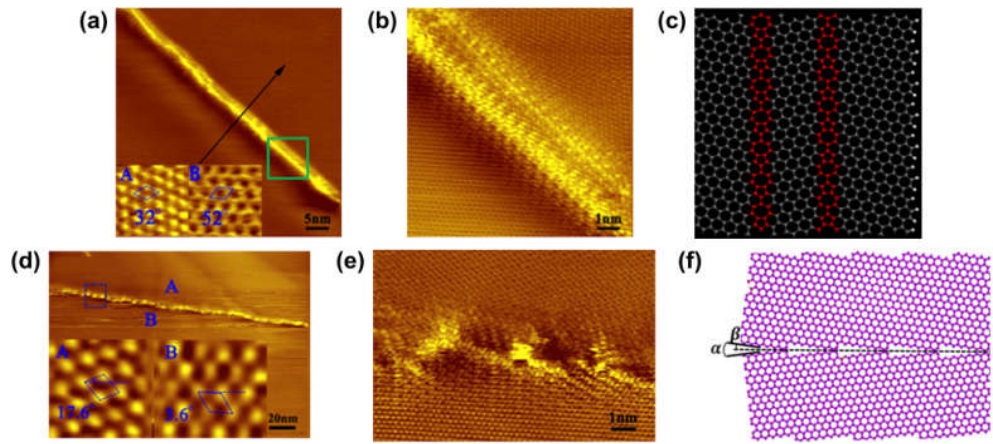

In this manuscript, the structures and electronic spectra of two ordered periodic grain boundaries (GBs), dual-parallel and zigzag-type GBs, on a graphene surface are studied using scanning tunneling microscopy and spectroscopy. Our results demonstrated that a GB consisting of dual parallel periodic dislocation cores of pentagonal–heptagonal (5–7) carbon rings results in an enhanced localized state at 0.45 eV above the Dirac point on graphene surfaces, which is attributed to van Hove singularities (VHSs). Moreover, the energy positions of the localized states are varied between 0.40 and 0.47 eV depending on the site and eventually decayed to 0.36 eV. The variation of the energy positions is induced by two parallel GBs because of the higher electron density at the GB as a defect center and the lower electron away from the GB. Another periodic GB with a symmetric zigzag structural feature induced VHSs at −0.038 and 0.12 eV near the Fermi level. We believe that this contribution is theoretically and practically relevant because it is the first study of a dual-parallel GB in graphene, and the periodic GBs embedded in graphene are promising with regard to improving the performance of graphene-based electronic devices.